\documentclass[aps,prb,twocolumn, groupedaddress,superscriptaddress,showpacs]{revtex4}
\usepackage{graphicx,amsmath}
\usepackage{amssymb}
\usepackage{nicefrac}

\begin{document}
\title{Semiclassical approximation solved by
Monte Carlo  as an efficient impurity solver for dynamical
mean field theory  and its cluster extensions}

\author{Hunpyo Lee}
\affiliation{Institut f\"ur Theoretische Physik, Goethe-Universit\"at Frankfurt, Max-von-Laue-Stra{\ss}e 1, 60438 Frankfurt am Main, Germany}
\affiliation{Division of Quantum Phases and Devices, School of Physics, Konkuk University, Seoul 143-701, Korea}
\author{Yu-Zhong Zhang}
\affiliation{Shanghai Key Laboratory of Special Artificial Microstructure Materials and Technology, School of Physics Science and engineering, Tongji University, Shanghai 200092, P.R. China}
\author{Hoonkyung Lee}
\affiliation{Division of Quantum Phases and Devices, School of Physics, Konkuk University, Seoul 143-701, Korea}
\author{Yongkyung Kwon}
\affiliation{Division of Quantum Phases and Devices, School of Physics, Konkuk University, Seoul 143-701, Korea}
\author{Harald O. Jeschke}
\affiliation{Institut f\"ur Theoretische Physik, Goethe-Universit\"at Frankfurt, Max-von-Laue-Stra{\ss}e 1, 60438 Frankfurt am Main, Germany}
\author{Roser Valent\'\i}
\affiliation{Institut f\"ur Theoretische Physik, Goethe-Universit\"at Frankfurt, Max-von-Laue-Stra{\ss}e 1, 60438 Frankfurt am Main, Germany}

\date{\today}

\begin{abstract}
  We propose that a combination of the semiclassical approximation
  with Monte Carlo simulations can be an efficient and reliable
  impurity solver for dynamical mean field theory equations and their
  cluster extensions with large cluster sizes. In order to show the
  reliability of the method, we consider two test cases: (i) the
  single-band Hubbard model within the dynamical cluster approximation
  with 4- and 8-site clusters and (ii) the anisotropic two-orbital
  Hubbard model with orbitals of different band width within the
  single-site dynamical mean field theory. We compare the critical
  interaction $U_c/t$ with those obtained from solving the dynamical
  mean field equations with continuous time and Hirsch-Fye quantum
  Monte Carlo.  In both test cases we observe reasonable values of the
  metal-insulator critical interaction strength $U_c/t$ and the nature
  of Mott physics in the self-energy behavior. While some details of the
  spectral functions cannot be captured by the semiclassical
  approximation due to the freezing of dynamical fluctuations, the
  main features are reproduced by the approach.

\end{abstract}



\maketitle

\section{Introduction\label{Introduction}}

The single-site dynamical mean field theory (DMFT) approach has
been extensively employed to explore the properties of the
Hubbard model and, in general, of strongly
correlated materials~\cite{Georges1996,Kotliar2004,Kotliar2006}.
Though the metal-Mott insulator transition can be
successfully accounted for within the DMFT approximation where
dynamical fluctuations alone are emphasized, interesting physical
phenomena such as spin density waves and superconductivity can
not be properly described due to the absence of spatial
fluctuations.  Cluster-extensions of
DMFT like cellular-dynamical mean field theory
(CDMFT)~\cite{Kotliar2006}
and dynamical cluster approximation
(DCA)~\cite{Hettler1998,Maier2005} can
take into account
the inter-site spatial fluctuations within the size of the
cluster in addition to the dynamical fluctuations.  For example, cluster
extensions of DMFT with 4-site clusters in combination with continuous
time quantum Monte Carlo (CT-QMC) or exact diagonalization (ED) can
partly capture the physics of Fermi liquid (FL), non-FL (or pseudo gap),
Mott insulator, and superconductivity on an equal
footing~\cite{Civelli2005,Zhang2007,Sakai2009,Liebsch2009,Sentef2011,Sordi2012,Tocchio2012}.
However, due to the computational expense of CT-QMC and ED, the system
size is still limited to only small clusters and not all physical
properties can be equally precisely studied.
In particular, the hybridization expansion CT-QMC approach is able to
treat  small cluster sizes  up to only $N_c=4$ due to an exponential increase
of the local Hilbert space with $N_c$~\cite{Werner2006(1),Werner2006(2)}.
The ED approach encounters a similar problem.
Even though other impurity solvers like   the
 interaction expansion CT-QMC approach are
applicable to large cluster systems, the computational expense is
proportional to the square of  three quantities: the interaction strength $U$, the
inverse of the temperature $T$, and the number of cluster sites
$N_c$~\cite{Rubtsov2005,Gull2011}. The Hirsch-Fye QMC (HF-QMC)~\cite{Hirsch1985,Hirsch1986} impurity solver
shows a computational expense proportional to $N_c^3L^3$ where $L$ is the number of slices in the
 imaginary time (temperature).
More recently, Khatami {\it et al.}~\cite{Khatami2010} proposed
the determinantal QMC (DQMC)~\cite{Blankenbecler1981} as a new impurity solver
where the computational expense has
a $(N_c+N_cN_a)^3L$ dependence with $N_a$ being the number of bath sites
connected to each cluster site. This is though a
Hamiltonian-based impurity solver that requires an explicit
form of  a cluster Anderson  impurity model to calculate the self-energy.
In contrast, CT-QMC, Hirsch-Fye QMC and  the method under discussion in the present
work, the semiclassical approximation (SCA),  are action-based impurity solvers.
Summarizing and in view of the above,
a fast and reliable impurity solver for DMFT
calculations and its cluster extensions with large cluster sizes is still highly desirable.

The SCA has been proposed as an impurity
solver for DMFT and its cluster
extensions~\cite{Okamoto2005,Fuhrmann2007,Lee2008}. (i) This impurity solver
is fast since the computational expense depends only on 
calculation time at each Matsubara frequency of the inverse of a matrix 
with dimensions $N_c \times N_c$ (for a cluster with $N_c$ sites) 
or $L_c \times L_c$ (for a single site with $L_c$ orbitals), 
where $L_c$ is the number of orbitals. (ii) While it cannot properly account for
Fermi-liquid behavior in the weak-coupling limit~\cite{Okamoto2005}, and in 
general, it is not adequate at  low temperatures due to the freezing
of quantum fluctuations in the method,
it is especially suited for  large interaction strength and multi-sites
where for example the powerful interaction expansion CT-QMC method is very costly.
 (iii) It provides self-energy information directly on the real frequency axis.
This avoids the uncertainty from
analytic continuation which has to be done in various QMC
approaches~\cite{Wang2009}. The previously used SCA
approach~\cite{Okamoto2005,Fuhrmann2007,Lee2008} was limited
though to consider small cluster sizes of $N_c=4$ due to the
difficulty of the multi-dimensional integrations.

In the present work, we propose to combine the SCA approach with
the Monte Carlo (MC) method. The latter is used to evaluate the
multi-dimensional integrals. We apply our scheme to two test cases: (i) the one-orbital
Hubbard model on the square lattice at half-filling  within the
DCA with  cluster sizes $N_c=4$ and $N_c=8$ and (ii)  the
anisotropic two-orbital Hubbard model with different band widths
on the Bethe lattice at half-filling within single-site DMFT.
We present the density of states, momentum dependent spectral
functions, and
momentum dependent self-energy as a function of real frequency
$\omega$.
For case (i) we find that even though
the Fermi liquid behavior is not obtained~\cite{comment1}, the
critical onsite Coulomb interaction $U_c/t$ for the
metal-insulator transition calculated by our SCA approach for both
cluster sizes shows a reasonable agreement with the value
obtained from CT-QMC
which should be numerically exact. In particular, we find that
the behavior of the density of states at the Fermi level in each
momentum sector obtained from both SCA and CT-QMC approaches are
quantitatively consistent with each other. For case (ii) we also
find a reasonable agreement of  $U_c/t$  obtained from
SCA and HF-QMC. The orbital-selective phase transition is also
correctly detected. However, we observe that if
the band-width difference  between
narrow and wide orbitals is large, a causality problem
appears in the SCA results.

The  paper is organized as follows. In
Sec.~\ref{Formalism} we present the general formalism of the
semiclassical approximation and its application to
cases (i) and (ii).
In Sec.~\ref{Results}, we discuss our SCA calculations and
compare some results with both CT-QMC and HF-QMC  and finally,
in Sec.~\ref{Conclusions} we summarize our findings.

\section{SEMICLASSICAL APPROXIMATION}\label{Formalism}

In this section we will review the formalism of the SCA
approach~\cite{Okamoto2005,Fuhrmann2007,Lee2008} adapted
to the two test cases considered in this work; the
8-site DCA and two-orbital DMFT systems with paramagnetic
solutions.

\subsection{General formalism}
The partition function can be written as:
\begin{equation}
Z = \int D \left[c^{\dagger} c \right] e^{-(S_{0} + S_{\rm int})},
\label{eq:partition}
\end{equation}
where
\begin{equation}
S_{\rm int}=U\int_0^{\beta} d\tau \sum_i n_{i\uparrow} (\tau) n_{i\downarrow} (\tau)
\end{equation}
 and
\begin{equation}
S_{0} = -\int_0^{\beta} d\tau \int_0^{\beta}d\tau' \psi_{\sigma}^{\dagger}(\tau) \hat{a}_{\sigma}(\tau, \tau')\psi_{\sigma} (\tau'),
\label{eq:action0}
\end{equation}
where $\psi_{\sigma}^{\dagger}=(c_{1\sigma}^{\dagger}\ldots
c_{l\sigma}^{\dagger})$ and $c_{i\sigma}^{\dagger}$ ($c_{i\sigma}$) is a
Grassmann number
corresponding to the Fermionic creation (annihilation) operator at site $i$
and spin $\sigma$, and $\hat{a}_{\sigma}=\sum_{\xi} a_{\xi\sigma}
\hat{K}_{\xi\sigma}$ where $a_{\xi\sigma}$ are inverted  frequency-
dependent Weiss fields and $\hat{K}_{\xi\sigma}$ are $l \times l$
matrices defined according to the chosen  cluster. Here, $l$ denotes the
number of sites in the multi-site system (case (i))  or two times the number of orbitals
in the multi-orbital system (case(ii)), and $\xi$ denotes the distance between two sites
within the cluster. For example, $a^{-1}_{0\sigma}$ means the local (on-site)
Weiss field while $a^{-1}_{\xi\sigma}$ is the inter-site Weiss field with the sites
located at a distance $\xi$-th apart. The orthogonality is imposed by
\begin{equation}
{\rm Tr} \left[\hat{K}_{\xi\sigma} \hat{K}_{\xi'\sigma'}\right] = l \delta_{\xi
\xi'}\delta_{\sigma \sigma'}.
\label{eq:normal}
\end{equation}
$n_{i\uparrow} (\tau) n_{i\downarrow} (\tau)$ in the
decoupling scheme can be written as:
\begin{equation}
n_{i\uparrow}(\tau)n_{i\downarrow}(\tau) = \frac{1}{4} (N_i(\tau)^2
 - M_i(\tau)^2),
\label{decoupling}
\end{equation}
where $N_i=(n_{i\uparrow}+n_{i\downarrow})$ and $M_i=(n_{i\uparrow}-
n_{i\downarrow})$ are the
particle number and magnetization,
respectively. In terms of these definitions and within the SCA approximation,
the partition function transforms into
\begin{equation}
Z = \int D \left[c^{\dagger} c \right] e^{\int_0^{\beta} d\tau \int_0^{\beta}d\tau'
\psi_{\sigma}^{\dagger}(\tau) \hat{a}_{\sigma}(\tau, \tau')\psi_{\sigma}
(\tau') +
\frac{U}{4}\int_0^{\beta} \sum_i M_i^{2} (\tau)}.
\label{eq6}
\end{equation}
where the $N_i^{2}(\tau)$ term, which describes  charge fluctuations, is
neglected in the SCA approximation.  This expression can be rewritten as
\begin{equation}
Z = \int D \left[c^{\dagger} c \right] \int_{-\infty}^{\infty} \prod_{i=1}^k d\phi_{i} e^{S},
\end{equation}
with
\begin{equation}
S = \int_0^{\beta}d\tau \bigg( \int_0^{\beta}
d\tau' \psi_{\sigma}^{\dagger} (\tau) \hat{a}_{\sigma} (\tau,
\tau')\psi_{\sigma} (\tau')
- \Big(\frac{\phi_i^2}{4U} - \frac{\phi_i M_i(\tau)}{2}\Big)\bigg).
\label{eq:action}
\end{equation}
Here we assume that the new auxiliary fields $\phi_i (\tau)$,
which are given by a continuous Hubbard-Stratonovich
transformation, are
$\tau$-independent $(\phi_i (\tau)\equiv \phi_i)$.

We replace $M_i(\tau)$ by
\begin{equation}
M_i(\tau)=\int_0^{\beta} d\tau' \sum_{ss'} c_{is}^{\dagger} (\tau) \sigma^{z} \delta (\tau - \tau') c_{is'} (\tau'),
\label{eq:mag}
\end{equation}
where $\sigma^z$ is the third Pauli matrix. Via a Grassmann
integration and Fourier transformation, the partition function
(Eq.~(\ref{eq6})) is finally given as
\begin{equation}
Z =\int_{-\infty}^{\infty} \prod_{i=1}^k d\phi_i e^{\displaystyle -\frac{\beta
\phi_i^2}{4U} + \sum_{\omega_n} {\rm ln} {\rm Det}\left[-\beta
(\hat{a}_{\sigma}(i\omega_n)+\frac{1}{2}\phi_i\sigma_z)\right]},
\label{eq10}
\end{equation}
where $\omega_n$ are Fermionic Matsubara frequencies and $k$ is the dimension
of integrations. The impurity Green's function can be obtained as
$\hat{G}^{\rm imp}_{\sigma}
(i\omega_n)=\sum_{\xi} G_{\xi\sigma}
(i\omega_n)\hat{K}_{\xi\sigma}$ where
$G_{\xi\sigma} (i\omega_n)$ is
\begin{equation}
G_{\xi\sigma} (i\omega_n) = \frac{1}{l} \frac{\partial {\rm ln} Z}{\partial
a_{\xi\sigma}(i\omega_n)}
\label{eq:green}
\end{equation}
where $l$ is the normalization factor which is given by
Eq. (\ref{eq:normal}). The Green's function on the real
frequency $\omega$
is also calculated by Eq.~(\ref{eq:green}) with replacement of
$\omega_n$
into $\omega +i\delta$. In our calculations we consider a
broadening factor $\delta=0.003$.
The integration in  $i=N_c \times m \times (2m-1)$ dimensions for
classical fields $\phi_i$ is evaluated by the MC approach, where
$N_c$ is the number of cluster sites, $m$ is the number of
orbitals, and the
weight function $W(\phi_i)$ for the MC simulations is given as
\begin{equation}
{\rm ln} W(\phi_i) = -\frac{\beta \phi_i^2}{4U} + \sum_{\omega_n} \rm{ln}
\rm{Det}\left[-\beta (\hat{a}_{\sigma}(i\omega_n)+\frac{1}{2}\phi_i\sigma_z)\right].
\label{eq:weight}
\end{equation}

\subsection{8-site dynamical cluster approximation}
\begin{figure}
\includegraphics[width=0.31\textwidth]{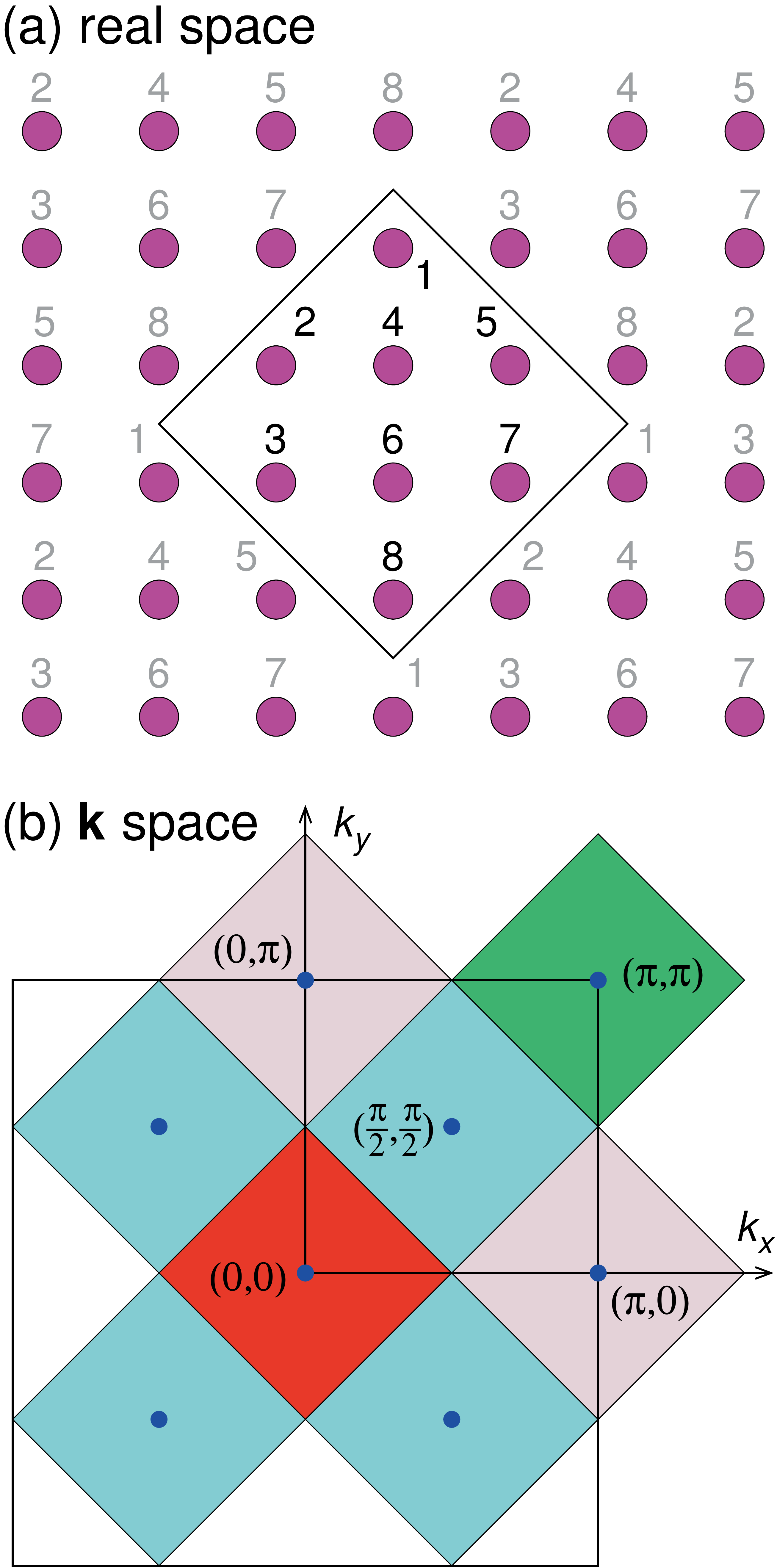}
\caption{(Color online) Cartoons for (a) the eight-site dynamical
  cluster approximation in real space and for (b) the division of the
  Brillouin zone in momentum space we used.}\label{Fig0:cluster}
\end{figure}

\label{eight-
site}
The partition function in the SCA approach is described in  a
real-space basis in Eqs.~\eqref{eq:partition}-\eqref{eq:action0}.
The $8 \times 8$ matrices of inversed Weiss fields (Eq.~(\ref{eq10}) and Eq.~(\ref{eq:green}))
 in the 8-site DCA calculations for the
Hubbard model on the square lattice are given as
\begin{equation}
\hat{a} (i\omega_n)=\left(\begin{array}{cccccccc}
a_{0}&\frac{a_{2}}{\sqrt{2}}&\frac{a_{1}}{2}&\frac{a_{1}}{2}&\frac{a_{2}}{\sqrt{2}}&a_{3}&\frac{a_{1}}{2}&\frac{a_{1}}{2}\\
\frac{a_{2}}{\sqrt{2}}&a_{0}&\frac{a_{1}}{2}&\frac{a_{1}}{2}&a_{3}&\frac{a_{2}}{\sqrt{2}}&\frac{a_{1}}{2}&\frac{a_{1}}{2}\\
\frac{a_{1}}{2}&\frac{a_{1}}{2}&a_{0}&\frac{a_{2}}{\sqrt{2}}&\frac{a_{1}}{2}&\frac{a_{1}}{2}&a_{3}&\frac{a_{2}}{\sqrt{2}}\\
\frac{a_{1}}{2}&\frac{a_{1}}{2}&\frac{a_{2}}{\sqrt{2}}&a_{0}&\frac{a_{1}}{2}&\frac{a_{1}}{2}&\frac{a_{2}}{\sqrt{2}}&a_{3}\\
\frac{a_{2}}{\sqrt{2}}&a_{3}&\frac{a_{1}}{2}&\frac{a_{1}}{2}&a_{0}&\frac{a_{2}}{\sqrt{2}}&\frac{a_{1}}{2}&\frac{a_{1}}{2}\\
a_{3}&\frac{a_{2}}{\sqrt{2}}&\frac{a_{1}}{2}&\frac{a_{1}}{2}&\frac{a_{2}}{\sqrt{2}}&a_{0}&\frac{a_{1}}{2}&\frac{a_{1}}{2}\\
\frac{a_{1}}{2}&\frac{a_{1}}{2}&a_{3}&\frac{a_{2}}{\sqrt{2}}&\frac{a_{1}}{2}&\frac{a_{1}}{2}&a_{0}&\frac{a_{2}}{\sqrt{2}}\\
\frac{a_{1}}{2}&\frac{a_{1}}{2}&\frac{a_{2}}{\sqrt{2}}&a_{3}&\frac{a_{1}}{2}&\frac{a_{1}}{2}&\frac{a_{2}}{\sqrt{2}}&a_{0}\\
\end{array}\right),
\label{eq:matrics}
\end{equation}
where spin indices are omitted for simplicity and the normalization
factors $\frac{1}{2}$ and $\frac{1}{\sqrt{2}}$ are introduced in order to
fulfill the orthogonality condition Eq.~(\eqref{eq:normal}). The indices
$\xi=0,1,2$, and $3$ indicate on-site, 1st neighbor, 2nd neighbor, and
3rd neighbor, respectively. The cluster we used for constructing the
$\hat{a} (i\omega_n)$ matrices with periodic boundary condition is shown
in  Fig.~\ref{Fig0:cluster} (a)  and the division of BZ for
DCA calculations is presented in
Fig.~\ref{Fig0:cluster} (b). The real-space impurity Green's functions in
Eq.~(\eqref{eq:green}) are more clearly expressed as
\begin{equation}
G_{\xi}(i\omega_n) = \frac{1}{l} \frac{1}{\rm{Det} \left[\hat{A}(a_{\xi}
(i\omega_n))\right]}
\frac{\partial}{\partial
a_{\xi}(i\omega_n)} \rm {Det} \left[\hat{A}(a_{\xi} (i\omega_n))\right],
\label{eq:impurity}
\end{equation}
where $\hat{A}(a_{\xi} (i\omega_n))=-\beta (\hat{a} (i\omega_n) +
\phi_i)$ are $8 \times 8$ matrices with spin index $\sigma$. 
The impurity Green's
function
in Eq. (\ref{eq:impurity}) is measured by
\begin{equation}
\frac{1}{\rm{Det} \left[\hat{A}(a_{\xi})\right]}\frac{\partial}{\partial a_{\xi}}
\rm{Det} \left[\hat{A}(a_{\xi})\right] = \rm{tr} \left[\hat{A}^{-1}(a_{\xi})
\frac{\partial \hat{A}(a_{\xi})}{\partial a_{\xi}}\right].
\end{equation}

\subsection{2-orbital dynamical mean field theory}
The interaction term  of the Hamiltonian for the 2-orbital system
(test case (ii)) is given as
\begin{equation}
H_{\rm int} = U\sum_{i\eta} n_{i\eta\uparrow}n_{i\eta\downarrow} + \sum_{i\sigma
\sigma'}(U'-\delta_{\sigma \sigma'} J_z)n_{i1\sigma}n_{i2\sigma'},
\label{eq:inter}
\end{equation}
where $\eta \in \{1,2\}$ denote orbital indexes. $U$ and $U'$
are, respectively, onsite
intra-orbital and inter-orbital Coulomb interaction parameters
and $J_z$ is the Ising Hund's coupling term. We are not
considering
the spin-flip and pair hopping terms in our calculations.
The inversed Weiss field is given as
\begin{equation}
\hat{a} (i\omega_n)=\left(\begin{array}{cccc}
a_{1,\uparrow} & 0 & 0 & 0\\
0 & a_{1,\downarrow} & 0 & 0\\
0 & 0 & a_{2,\uparrow} & 0\\
0 & 0 & 0 & a_{2,\downarrow}\\
\end{array}\right).
\end{equation}
We now decouple the interaction term Eq.~(\ref{eq:inter}) using
Eq.~(\ref{decoupling}):
\begin{eqnarray}
n_{1\uparrow}n_{1\downarrow} = \frac{1}{4} (N_{1}^2 - M_{1}^2),\nonumber
n_{2\uparrow}n_{2\downarrow} = \frac{1}{4} (N_{2}^2 - M_{2}^2),\\
n_{1\uparrow}n_{2\downarrow} = \frac{1}{4} (N_{3}^2 - M_{3}^2),\nonumber
n_{1\downarrow}n_{2\uparrow} = \frac{1}{4} (N_{4}^2 - M_{4}^2),\\
n_{1\uparrow}n_{2\uparrow} = \frac{1}{4} (N_{5}^2 - M_{5}^2),\nonumber
n_{1\downarrow}n_{2\downarrow} = \frac{1}{4} (N_{6}^2 - M_{6}^2).
\end{eqnarray}
Neglecting charge fluctuations $N_{\xi}^2$, the partition function can
be written  as
\begin{equation}
\frac{Z}{Z_{0}}= e^{\displaystyle \int_0^{\beta} (\frac{U}{4} (M_1^2 + M_2^2)
+ \frac{U'}{4} (M_3^2 + M_4^2) + \frac{U''}{4} (M_5^2 + M_6^2))},
\label{eq:parti}
\end{equation}
where $U'' = U'-J_z$ and
\begin{equation}
Z_0 = \int D \left[c^{\dagger} c \right] e^{\displaystyle \int_0^{\beta} d\tau
\int_0^{\beta} d\tau' \psi^{\dagger} (\tau) \hat{a}(\tau , \tau')
\psi(\tau')},
\end{equation}
where
$\psi^{\dagger}=(c^{\dagger}_{1\uparrow},c^{\dagger}_{1\downarrow},c^{\dagger}_{2\uparrow},c^{\dagger}_{2\downarrow})$
(compare with Eq.~(\ref{eq:action0})).
In order to make integration feasible, Eq. (\ref{eq:parti}) is transformed
into $Z = Z_{0} e^{-S}$ with
\begin{equation}\begin{split}
S= &-\frac{\phi_1^2 + \phi_2^2}{4U} - \frac{\phi_3^2 + \phi_4^2}{4U'} - \frac{\phi_5^2 + \phi_6^2}
{4U''}\\
&+ \int_0^{\beta}d\tau
\frac{1}{2}\sum_{\xi=1}^6 \phi_{\xi} M_{\xi} (\tau)
\end{split}\end{equation}
where we used  the continuous Hubbard-Stratonovich transformation
as in Eq. (\ref{eq:action}).
Next (see Eq.~(\ref{eq:mag})), $M_{\xi} (\tau)$ is replaced by
\begin{equation}
M_{\xi} (\tau)=\int_0^{\beta} d\tau' \psi^{\dagger} (\tau) \sigma_{\xi}^z \delta (\tau - \tau') \psi (\tau').
\end{equation}
Site indices are omitted due to the single-site DMFT calculation
and $\sigma_{\xi}^z$ are $4 \times 4$ matrices:
\begin{eqnarray}
\sigma_1^z = \left(\begin{array}{cccc}
1 & 0 & 0 & 0\\
0 & -1 & 0 & 0\\
0 & 0 & 0 & 0\\
0 & 0 & 0 & 0\\
\end{array}\right),\nonumber
\sigma_2^z = \left(\begin{array}{cccc}
0 & 0 & 0 & 0\\
0 & 0 & 0 & 0\\
0 & 0 & 1 & 0\\
0 & 0 & 0 & -1\\
\end{array}\right)\\
\sigma_3^z = \left(\begin{array}{cccc}
0 & 0 & 0 & 1\\
0 & 0 & 0 & 0\\
0 & 0 & 0 & 0\\
-1 & 0 & 0 & 0\\
\end{array}\right),\nonumber
\sigma_4^z = \left(\begin{array}{cccc}
0 & 0 & 0 & 0\\
0 & 0 & 1 & 0\\
0 & -1 & 0 & 0\\
0 & 0 & 0 & 0\\
\end{array}\right)\\
\sigma_5^z = \left(\begin{array}{cccc}
0 & 0 & 1 & 0\\
0 & 0 & 0 & 0\\
-1 & 0 & 0 & 0\\
0 & 0 & 0 & 0\\
\end{array}\right),\nonumber
\sigma_6^z = \left(\begin{array}{cccc}
0 & 0 & 0 & 0\\
0 & 0 & 0 & 1\\
0 & 0 & 0 & 0\\
0 & -1 & 0 & 0\\
\end{array}\right)
\end{eqnarray}
Finally, the partition function is rewritten as
\begin{equation}
Z = \int_{-\infty}^{\infty} \prod_{\xi=1}^6 d\phi_{\xi} e^{-V(\phi_{\xi})
+ \sum_{\omega_n} \rm{ln} \rm{Det}\left[-\beta (\hat{a}(i\omega_n)+\frac{1}
{2}\phi_{\xi}\sigma_{\xi}^z)\right]},
\end{equation}
where $V(\phi_{\xi}) = \beta (\frac{\phi_1^2 + \phi_2^2}{4U} + \frac{\phi_3^2 + 
\phi_4^2}{4U'}+\frac{\phi_5^2 + \phi_6^2}{4U''})$. The impurity Green's 
functions are calculated by Eq.~(\ref{eq:green}).

\subsection{Monte Carlo measurement}

The weight functions for MC calculations have been given in
Eq.~(\ref{eq:weight}).
We employ about 400 Matsubara frequencies for
performing the frequency sum in Eq.~(\ref{eq:weight}).
The number of classical fields $\phi_i$ is the same as the number of
cluster sites in the DCA calculations (case (i)).
For case (ii) the number of classical fields
is given by $(2m-1) \times m$,
where $m$ is the number of orbitals. In order to avoid a local
minimum problem in the MC calculation, we use around forty initial
configurations and we perform about $4 \times 10^{5}$ MC samplings for
each different initial configuration. The computational cost for $4 \times 10^{5}$ MC samplings
in the 8-site DCA system is around 90 minutes on a single 4~GHz CPU machine
and the error is smaller than
$5 \times 10^{-4}$.

\section{RESULTS}\label{Results}

\subsection{The metal-insulator transition in the 8-site dynamical cluster approximation}

In what follows, we will show the reliability of our SCA impurity
solver by presenting the results obtained for 4- and 8-site DCA
calculations for a two-dimensional Hubbard model on the square lattice
at half-filling (case (i)).

\begin{figure}
\includegraphics[angle=270,width=0.47\textwidth]{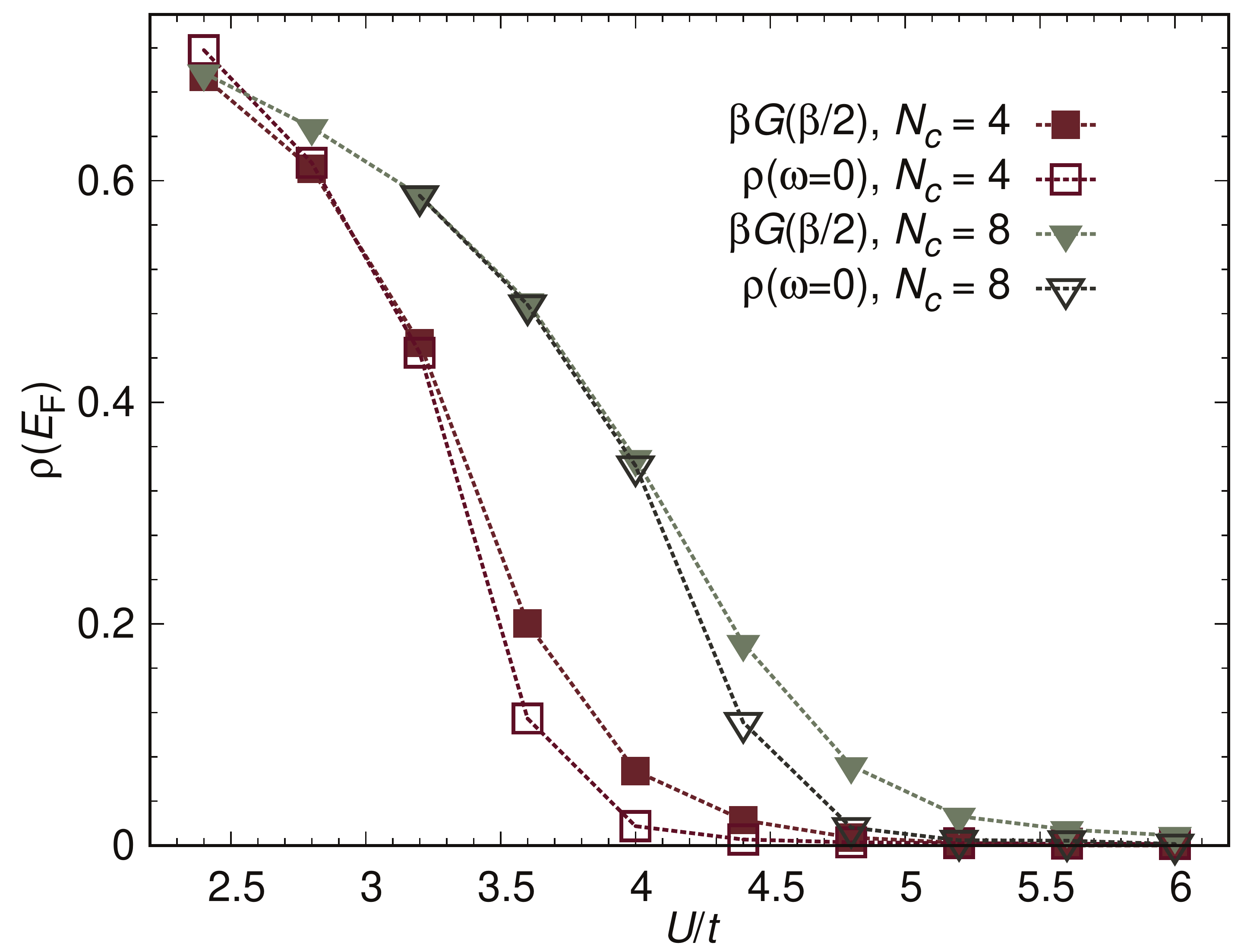}
\caption{(Color online) The density of states at the Fermi level $\rho
(\omega = 0) \approx \beta G(\frac{\beta}{2})$ and $\rho(\omega = 0)$,
directly measured in real-frequency space within DCA(SCA), as a function of
$U/t$ for $N_c=4$ and 8 at $T/t=1/12$.  From $\rho(\omega = 0)$
we find the critical metal-insulator
interactions $U_c/t= 4.4 \pm 0.2$ and $5.0 \pm 0.2$ for $N_c=4$ and 8, respectively. The error
bars are smaller than symbol sizes. The deviations of between $\beta
G(\frac{\beta}{2})$ and $\rho(\omega =0)$ are around ten
percent. See the main text for discussion.}\label{Fig1:DOS}
\end{figure}

First, we compare the critical $U_c/t$ obtained from DCA(SCA)
calculations with that obtained from DCA(CT-QMC) with $N_c = 4$ and
$8$.  Within DCA(CT-QMC)~\cite{Gull2008,Gull2009} $U_c/t=4.5$
($N_c=4$) and $U_c/t=6.5$ ($N_c=8$). We would like to note that for
the two-dimensional Hubbard model on the square lattice at
half-filling, previous DCA calculations showed that, at finite
temperature, the larger the cluster size is, the smaller the critical
value of interaction $U_c/t$. This is due to the fact that DCA
calculations account for spatial correlations only within the
cluster~\cite{Moukouri2001, Kyung2003}.  This suggests that we should
expect a smaller $U_c/t$ for increasing cluster sizes. However, this
is not what it is observed above. We think that the reason for this
discrepancy lies on the fact that plaquette singlet ordered states
become more favorable for $N_c=4$ than for $N_c=8$ and artificially
stabilize an insulating state in $N_c=4$. We now check whether this
behavior is also observed in the SCA approach.  In Fig.~\ref{Fig1:DOS}
we plot the DCA(SCA) density of states at the Fermi level obtained as
(1) $\rho (\omega = 0) \approx \beta G(\frac{\beta}{2})$~\cite{comment} and (2)
directly calculated in real-frequency space $\rho (\omega =0)$, as a
function of $U/t$ for $N_c=4$ and $8$ at $T/t =1/12$. The imaginary
time Green's function $G(\tau)$ is calculated by the Fourier
transformation of $G(i\omega_n)$ in Eq.~(\ref{eq:green}).  We find
$U_c/t = 4.4 \pm 0.2$ and $5.0 \pm 0.2$ for $N_c = 4$ and $8$ respectively. The trend,
{\it i.e.}, smaller critical interaction $U_c/t$ for $N_c=4$ than for
$N_c=8$, is the same as in DCA(CT-QMC).  We also detect that the
critical interactions $U_c/t$ in the SCA method are slightly smaller than those
calculated by the CT-QMC method. The reason why the insulating state
is overestimated, is due to the fact that the auxiliary field is
assumed to be $\tau$ independent in SCA, indicating a freezing of
dynamical fluctuations in SCA.

Usually one always employs the relation $\rho (\omega = 0) \approx
\beta G(\frac{\beta}{2})$ to determine the critical interaction
$U_c/t$ for the metal-insulator transition in
DCA(CT-QMC)~\cite{Gull2008,Gull2009}.  This is done in order to avoid
the performance of an analytical continuation, which will introduce
some uncertainties. Therefore, it is interesting to check whether this
relation is valid in all cases. Since DCA(SCA) provides results
directly in real-frequency space, both definitions can be tested on
the same footing. Fig.~ \ref{Fig1:DOS} shows that both definitions are
in good agreement in the weak-coupling and strong-coupling regions,
but they show deviations of about ten percent close to the critical
value due to finite temperature effects~\cite{comment}.
  In addition, we find a causality problem in the weak-coupling
region (for $U/t$ values smaller than $3.0$).

\begin{figure}
\includegraphics[angle=0,width=0.47\textwidth]{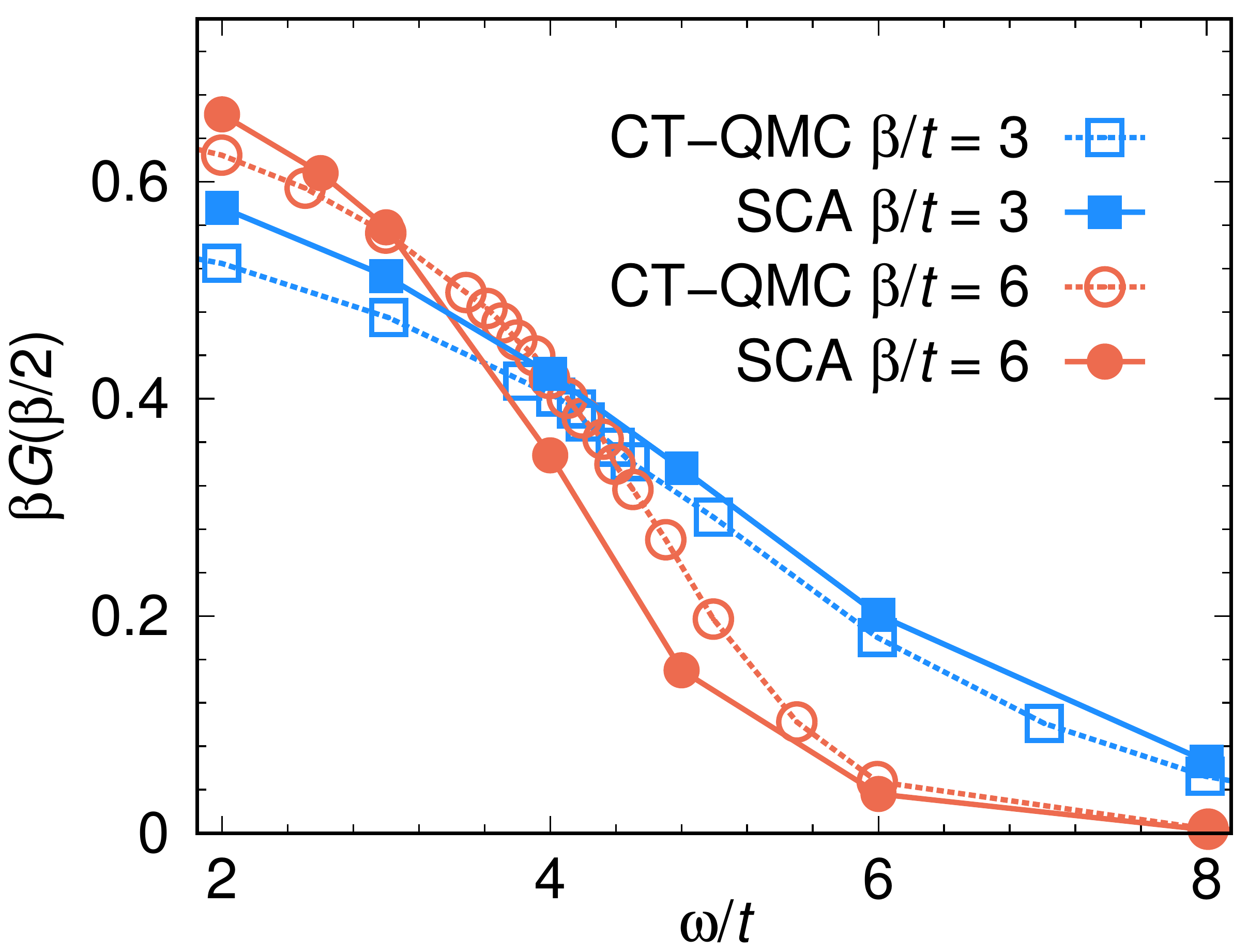}
\caption{(Color online)  Comparison of $\beta G(\frac{\beta}{2})$ in DCA(SCA) and
DCA(CT-QMC) for $\beta=3.0/t$ and $6.0/t$
 as a function of $\omega/t$
for  $N_c=4$. The CT-QMC results were obtained from  Ref.~\protect\onlinecite{Gull2008}.}
\label{density_comp}
\end{figure}

In order to test the reliability of the SCA approximation we present in Fig.~\ref{density_comp}
a quantitative comparison of $\beta G(\beta/2)$ as a function of frequency
obtained by the SCA impurity solver
and by CT-QMC~\cite{Gull2008} for $N_c=4$ and  inverse temperatures $\beta=3/t, 6/t$.
We observe a good agreement between both sets of results at high temperature regions.

\begin{figure}
\includegraphics[angle=0,width=0.47\textwidth]{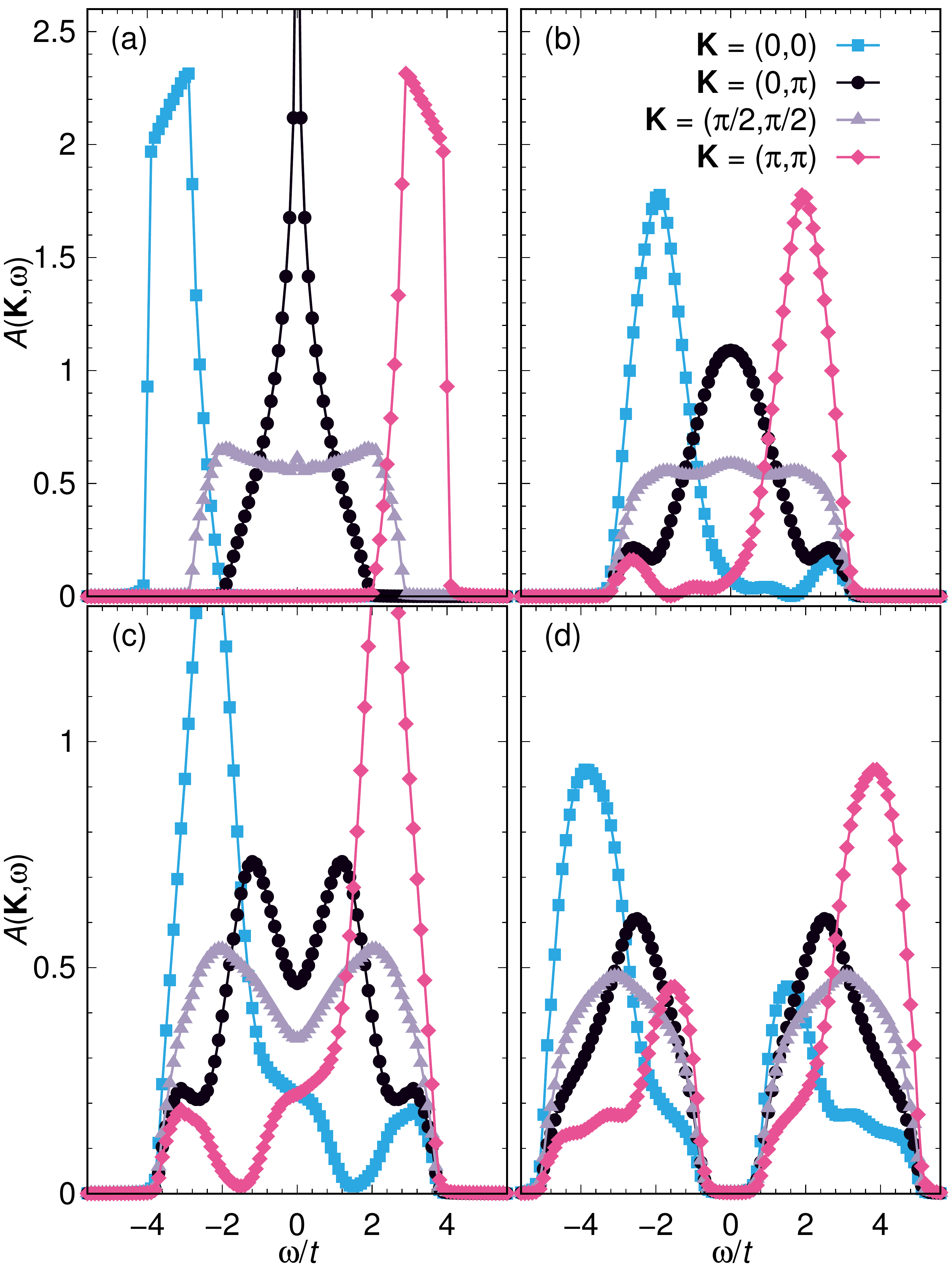}
\caption{(Color online) The spectral functions $A({\bf K},
\omega)$ in the different DCA momentum sectors ${\bf K}$ for (a)
$U/t=0.0$, (b) 3.2 (c) 4.0 and (d) 6.0 at $T/t=1/12$.}\label{Fig2:SP}
\end{figure}

Next, we would like to analyze the spectral functions in different DCA
cluster momentum sectors, {\it i.e.}, $A({\bf K},\omega)$ at
${\bf K}=(0,0)$, $(0,\pi)$, $(\frac{\pi}{2},\frac{\pi}{2})$ and
$(\pi ,\pi)$
shown in Fig.~\ref{Fig0:cluster}~(b) for several values of
$U/t$ at $T/t=1/12$.
In  Fig.~\ref{Fig2:SP}~(a) we display  the
non-interacting case ($U/t=0.0$).  While the weights of the
spectral functions $A({\bf K},\omega)$ at ${\bf K}=(0,0)$ and
$(\pi,\pi)$ sectors are well separated from each other, resembling the
behavior of  band insulators, the spectral functions at
${\bf K}=(0,\pi)$ and $(\frac{\pi}{2},\frac{\pi}{2})$ sectors cross the
Fermi level, showing metallic behavior. The van-Hove
singularity is present in the spectral function in the
${\bf K}=(\pi,0)/(0,\pi)$ sector.  The behavior of
$A({\bf K},\omega)$ for $N_c=8$ is comparable to  $N_c=4$
results~\cite{Haule2007} and can be understood in terms of the non-interacting band structure.

\begin{figure}
\includegraphics[width=0.47\textwidth]{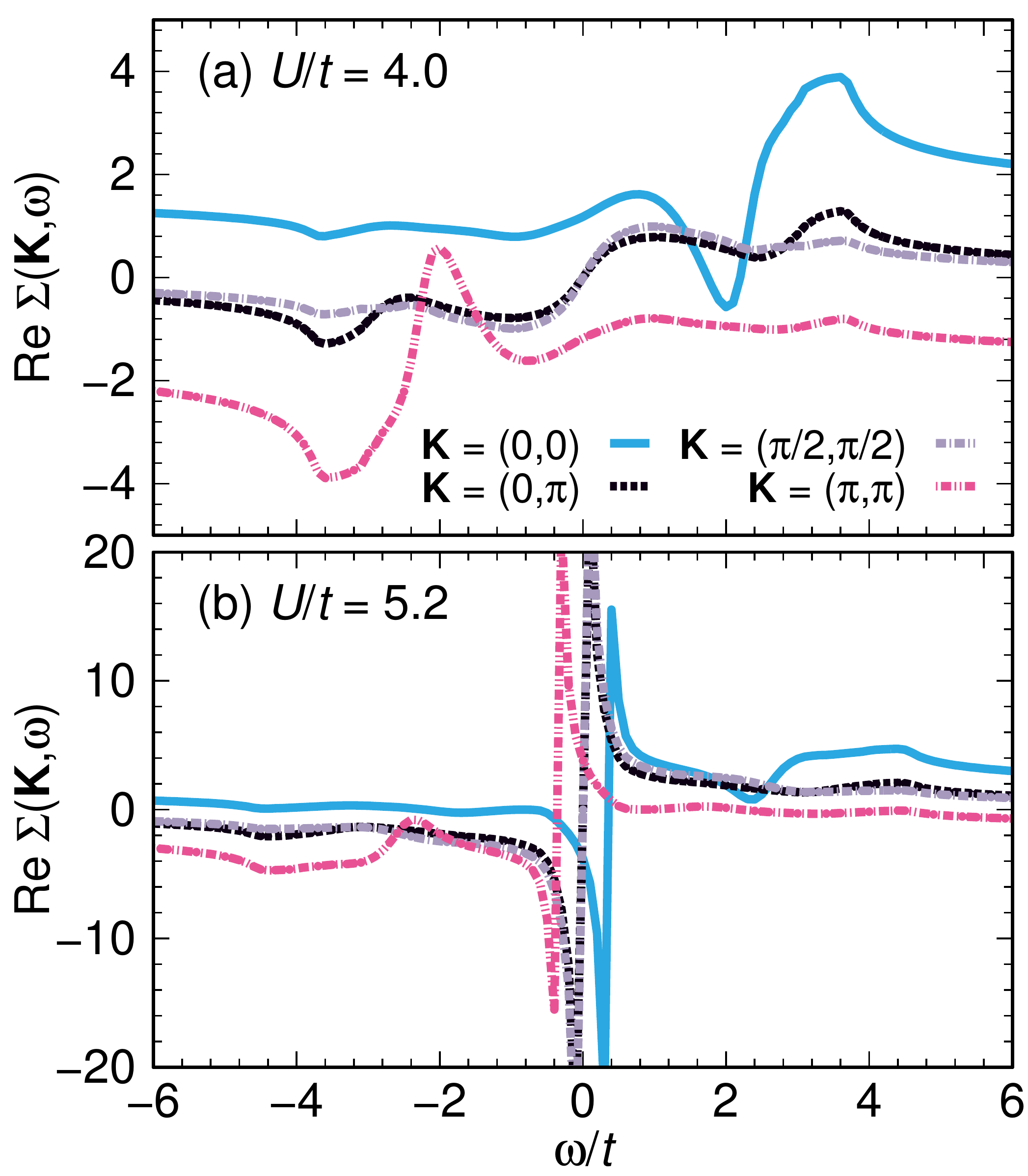}
\caption{(Color online) The real part of self-energy ${\rm Re}
(\Sigma({\bf K},\omega))$ as a function of
real frequency $\omega /t$ for (a) $U/t=4.0$ and (b) $U/t=5.2$ at
$T/t=1/12$.}\label{Fig3:selfenergy}
\end{figure}

In Fig.~\ref{Fig2:SP}~(b) we show $A({\bf K},\omega)$ for $U/t=3.2$
(weak-coupling region). $A({\bf K},\omega)$ at ${\bf K}=(0,0)$ and
$(\pi,\pi)$ intersect with each other due to the band splitting and
spectral weight transfer induced by $U$.  This is an indication of
Mott physics. The insulating behavior still remains in these two
sectors and the whole band width is slightly narrowed due to
correlation effects. The van-Hove singularity that was present in the
${\bf K}=(\pi,0)$ sector in the non-interacting case, is dramatically
suppressed with increasing $U/t$. The absence of a strong
quasi-particle peak in the weak interaction region is due to the
freezing of dynamical fluctuations. This is a shortcoming of the SCA
method.  As the interaction $U/t$ is increased, a pseudo-gap behavior
is present with suppression of the spectral functions at ${\bf
  K}=(0,\pi)$ and $(\frac{\pi}{2},\frac{\pi}{2})$ at $U/t=4.0$
(Fig.~\ref{Fig2:SP}~(c)), and the Mott insulator appears in the
strong-coupling region at $U/t=6.0$ (Fig.~\ref{Fig2:SP}~(d)).

\begin{figure}
\includegraphics[angle=270,width=0.47\textwidth]{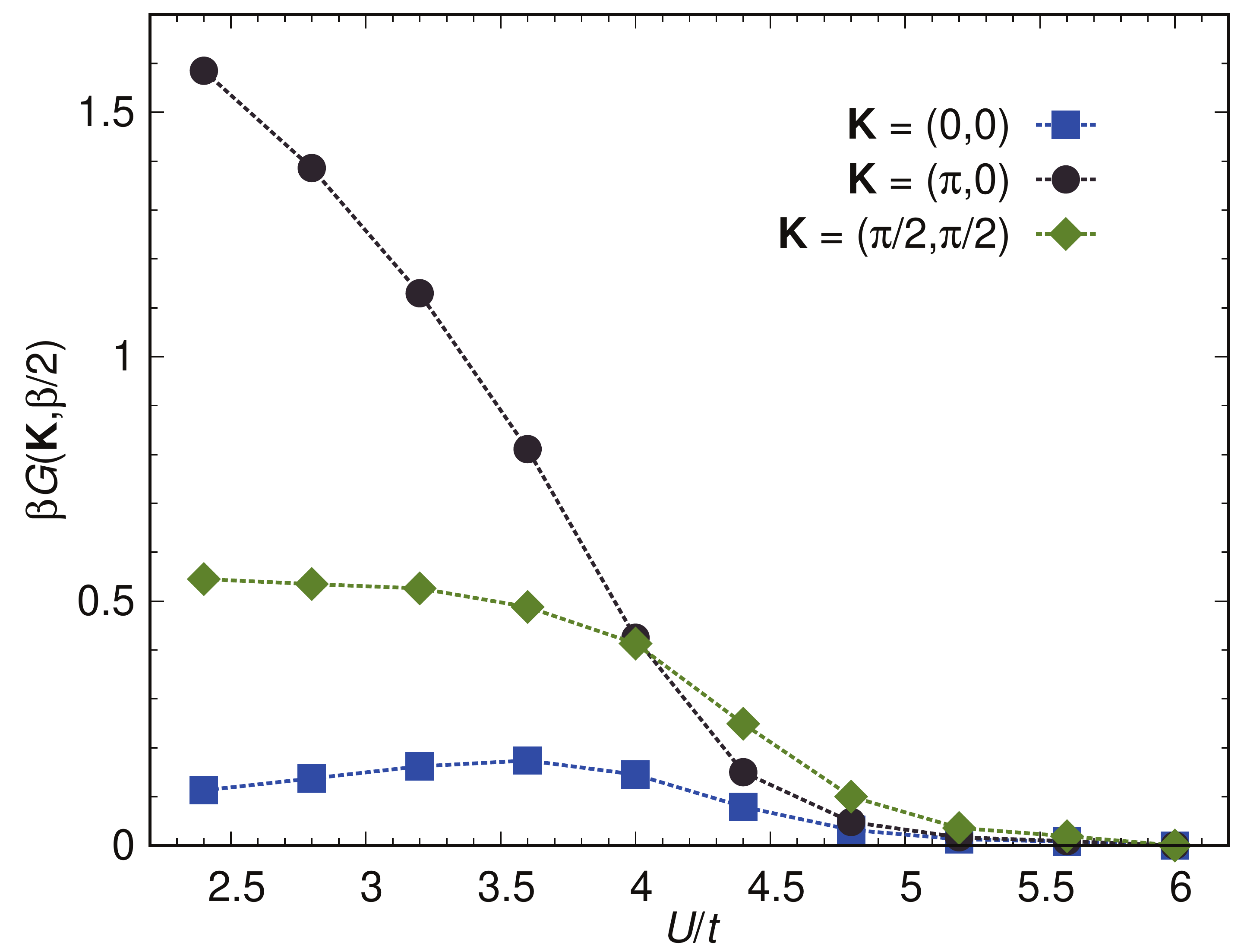}
\caption{(Color online) The quantity $\beta G({\bf K},
  \frac{\beta}{2})$ in the different DCA momentum sectors ${\bf K}$ in
  the 8-site DCA approach at $T/t=1/12$.  The values of $\beta G({\bf
    K}, \frac{\beta}{2})$ at ${\bf K}=(\pi, \pi)$ are the same as
  those at ${\bf K}=(0, 0)$.  The error bars are smaller than symbol
  sizes. These results are comparable with those obtained from the
  interaction-expansion CT-QMC approach in
  Ref.~\onlinecite{Gull2009}.}\label{Fig3:SPT}
\end{figure}

In order to study the Mott behavior in 8-site DCA in more detail, we
present in Fig.~\ref{Fig3:selfenergy} the real part of the self-energy
${\rm Re} (\Sigma({\bf K},\omega))$ as a function of real frequency
$\omega /t$ for $U/t=4.0$ and $5.2$ at $T/t=1/12$.  The real and
imaginary parts of the self-energy give the energy shift and the
spectral broadening, respectively, of the one-electron spectrum due to the
interaction $U/t$.
In Fig.~\ref{Fig3:selfenergy}~(a) for $U/t=4.0$ where the spectral
function shows a pseudo-gap, the real parts of the self-energy ${\rm
  Re} (\Sigma({\bf K},\omega))$ at ${\bf K}=(0,0)$ and $(\pi,\pi)$
remain finite below and above the Fermi level, respectively,
indicating the shift of the pole positions of the one-electron spectrum.
At ${\bf K}=(0,\pi)$  and $(\pi /2 , \pi /2)$  ${\rm
  Re} (\Sigma({\bf K},\omega))$  shows a positive slope with negative value of quasiparticle weight and the corresponding
${\rm Im} (\Sigma({\bf K},\omega))$ (not shown) exhibits a peak around the Fermi level,
indicating the appearance of a non-Fermi liquid. This non-Fermi-liquid behavior is
the sign of the Mott gap beginning to form.
In Fig.~\ref{Fig3:selfenergy} (b) the $U/t=5.2$ case is presented.
It is known that, in the Mott insulating state,
the self-energy has a pole-like structure of the form
\begin{equation}
\Sigma(\omega) \propto \frac{1}{\omega - \triangle + i\gamma},
\label{equation:pole}
\end{equation}
where the damping $\gamma$ is small, and $\triangle$ is the position
of the pole~\cite{Lin2009}.
We observe that while the ${\rm Re} (\Sigma({\bf K},\omega))$ at ${\bf K}=(0,\pi)$  and $(\pi /2 , \pi /2)$
for $\omega \rightarrow 0$ shows a pole-like structure indicating the Mott insulating state, the pole in the ${\rm Re} (\Sigma({\bf K},\omega))$ at ${\bf K}=(0,0)$ and $(\pi,\pi)$ lies above and below Fermi level, respectively.

In Fig.~\ref{Fig3:SPT} we show $\beta G({\bf K}, \frac{\beta}{2})$ as
a function of $U/t$ at $T/t=1/12$.  When the interaction $U/t$ is
turned on, the values of $\beta G({\bf K}, \frac{\beta}{2})$ at ${\bf
  K}=(0,0)$ increase until $U/t=3.6$ due to spectral weight transfer
caused by electronic correlations, and a metallic behavior is seen in
the ${\bf K}=(0,0)$ sector in the intermediate interaction strength
regions.  When the interaction becomes strong, the gap opens and
$\beta G({\bf K}, \frac{\beta}{2})$ goes to zero.  In the ${\bf
  K}=(0,\pi)$ sector $\beta G({\bf K}, \frac{\beta}{2})$ decreases
monotonously and goes to zero at $U/t=5.6$.  In the ${\bf
  K}=(\frac{\pi}{2},\frac{\pi}{2})$ sector, the values of $\beta
G({\bf K},\frac{\beta}{2})$ remain nearly constant up to
$U/t=3.0$. Beyond $U/t=3.0$, they decrease and the gap opens
completely around $U/t=5.6$.

Finally, we compare the results in Fig.~\ref{Fig3:SPT} to those in
Fig. 8 of Ref.~\onlinecite{Gull2009} calculated within the CT-QMC
approach. The behavior at ${\bf K}=(0,\pi)$ in both SCA and CT-QMC
approaches is qualitatively the same, even though the critical
interactions $U_c/t$ are different. The main difference between these
two results is in the ${\bf K}=(\frac{\pi}{2},\frac{\pi}{2})$
sector. The CT-QMC results indicate a first-order transition with a
discontinuous behavior of $\beta G({\bf K},\frac{\beta}{2})$ at the
critical interaction $U_c/t$, while the SCA results show a continuous
transition with a smooth decrease of $\beta G({\bf
  K},\frac{\beta}{2})$. We think that this discrepancy between the two
approaches also comes from the approximation that the dynamical
fluctuations are frozen in the SCA method.

\subsection{Orbital-selective phase transitions in the two-
orbital dynamical mean field theory}

\begin{figure*}
\includegraphics[width=0.9\textwidth]{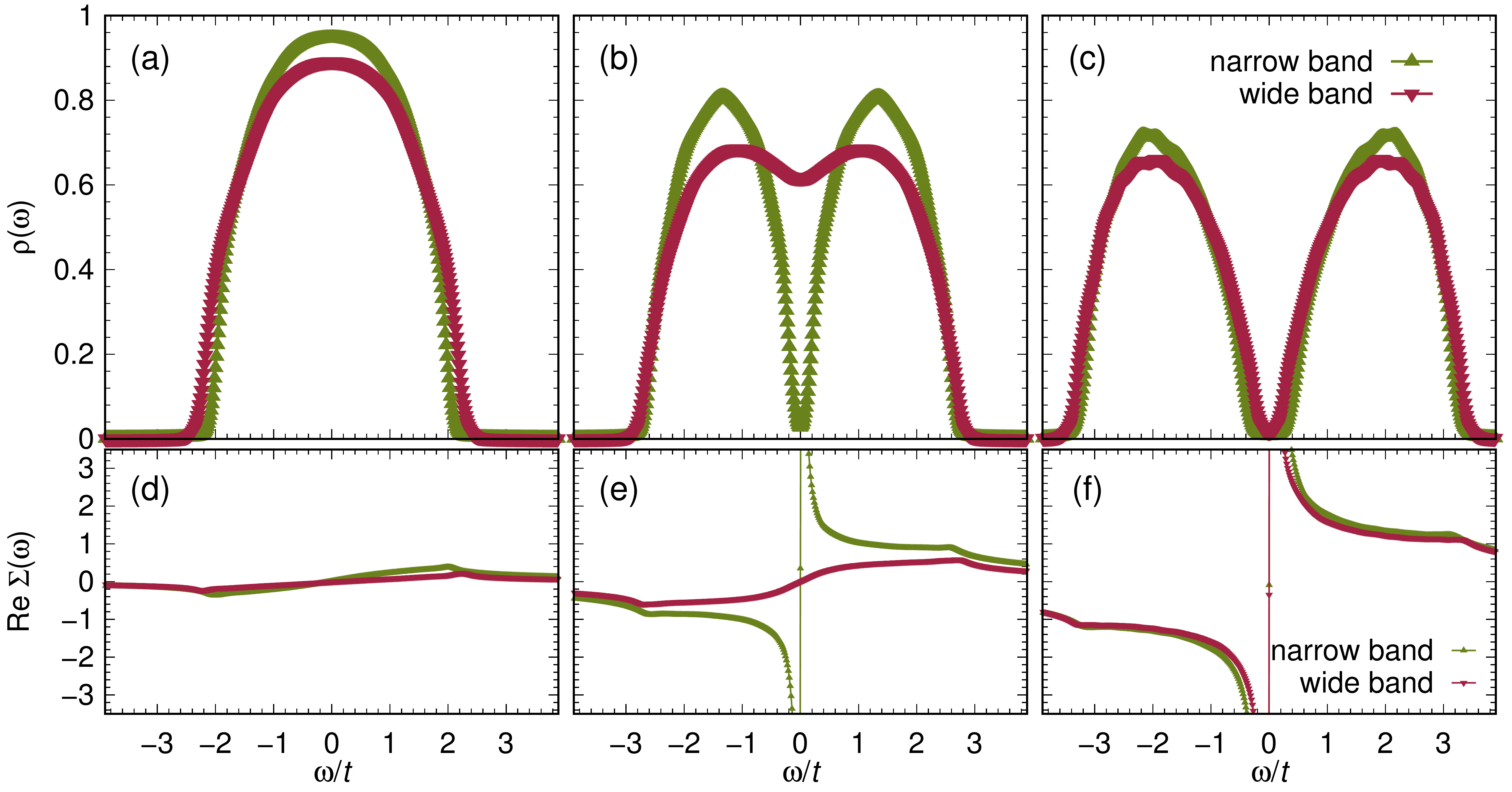}
\caption{(Color online) The density of states $\rho(\omega)$ and
real part of the self-energy ${\rm Re} (\Sigma(\omega))$ for
(a) and (d) $U/t_2=2.0$, (b) and (e) 2.8, and (c) and (f) 3.6 at
$T/t_2=1/12.0$, respectively. The
bandwidths for narrow and wide orbitals on the Bethe lattice are
$W_1=2.56$ ($t_1=0.8$) and $W_2=4.0$ ($t_2=1.0$), respectively.
The metal, orbital-selective phase, and Mott insulator are
present in (a), (b), and (c), respectively.}\label{Fig4:DOS}
\end{figure*}

The orbital-selective phase transition (OSPT), where metallic behavior
is seen in the wide band while a metal-insulator transition is
observed in narrow band, has been intensively studied in model systems
as well as real materials during the last ten
years~\cite{Anisimov2002,Koga2004,Knecht2005,Liebsch2005,Inaba2005,
  Biermann2005,Bouadim2009,Lee2010,Lee2011,Lee2012}.  We study the
anisotropic two-orbital Hubbard model with a narrow bandwidth of
$W_1=2.56$ ($t_1=0.8$) for the first orbital and a wide bandwidth of
$W_2=4$ ($t_2 = 1.0$) for the second orbital at half-filling on the
Bethe lattice using single-site DMFT. The interaction part of the
Hamiltonian is given by Eq.~(\ref{eq:inter}) with $U'=\frac{U}{2}$ and
$J_z=\frac{U}{4}$.  The density of states $\rho (\omega)$ in both orbitals obtained with
the SCA are shown in Fig.~\ref{Fig4:DOS}~(a)-(c).  Metallic behavior
in both bands is observed for the weak coupling strength $U/t_2=2.0$
in Fig.~\ref{Fig4:DOS}~(a). As the interaction $U/t$  increases, an orbital
selective phase transition behavior is present in the intermediate regions for $U/t_2=2.8$
(see Fig.~\ref{Fig4:DOS}~(b)).  Finally, insulating states in both
orbitals are seen in the strong-coupling region for $U/t_2=3.6$ in
Fig.~\ref{Fig4:DOS}~(c). We also present the real part of the
self-energy ${\rm Re} (\Sigma(\omega))$ as a function of real
frequency $\omega/t$ in Figs.~\ref{Fig4:DOS}~(d)-(f). As discussed for
the 8-site DCA results, the Mott insulating state is related to a
polelike structure in the self-energy as in
Eq.~\eqref{equation:pole}.  For $U/t_2=2.0$
 ${\rm Re} (\Sigma(\omega))$ in both orbitals is small
(Fig.~\ref{Fig4:DOS}~(d)). Increasing $U/t$ to $2.8$ we observe that
${\rm Re} (\Sigma(\omega))$ in the narrow-band orbital becomes large near the Fermi level
(Fig.~\ref{Fig4:DOS}~(e)) while it retains its small value for the wide-band orbital (OSPT region).
Finally  for $U/t_2=3.6$,  both
self-energies show poles at $\omega/t=0$, indicating the Mott insulating behavior in both orbitals.

In the following we compare the critical values of the interaction
strength obtained from the SCA and HF-QMC calculations in the
case of band widths of $W_1=2$ (narrow band) and $W_2=4$ (wide band).
 The
critical values in the HF-QMC approach are given as
$U_{c_1}/t_2=2.0$ and $U_{c_2}/t_2=2.8$ in narrow and wide
bands, respectively~\cite{Knecht2005}.  Our SCA results show the
critical
values $U_{c_1}/t_2=2.0$ and $U_{c_2}/t_2=3.2$ from the analysis
of the Green's function in the Matsubara frequency space
in good agreement with HF-QMC.  We
encounter though a
causality problem if the difference between the band widths of
narrow and wide orbitals is large.

\section{SUMMARY}\label{Conclusions}

In summary, in this work we propose that the semiclassical
approximation in combination with the Monte Carlo method can be used
to study large clusters and multi-orbital systems and is easy to embed
into DMFT and its cluster extensions.  We investigate the
single-orbital Hubbard model by the DCA(SCA) method with cluster sizes
of $N_c=4$ and 8 and a two-orbital system by DMFT(SCA).  The critical $U_c/t$
as well as $G(\frac{\beta}{2})$ as a function of $\omega$
are compared with existing DCA(CT-QMC)  results. The
critical interactions $U/t$ of SCA and CT-QMC approaches in both cases
are in reasonable agreement.  In the 8-site DCA cluster calculation,
we analyze the spectral functions $A({\bf K},\omega)$ and self-energy at each momentum
sector.  The only difference between SCA and CT-QMC results is that
the CT-QMC shows a discontinuous behavior of the spectral density in
the ${\bf K}=(\frac{\pi}{2},\frac{\pi}{2})$ sector around the critical
value of interactions, while the SCA exhibits smoothly decreasing
behavior. We think that the reason for the discrepancy is that quantum
fluctuations are frozen in the SCA approach.  In the two-orbital
DMFT(SCA) calculation, we observe the orbital selective phase
transition as in previous studies performed with
DMFT(HF-QMC)~\cite{Knecht2005}. This method is rather powerful
since it can be applied for problems where other impurity solvers
remain computationally too expensive but one should be aware of
possible causality problems in some cases.

\section{ACKNOWLEDGMENTS}
We would like to thank Gang Li, Hartmut Monien and Claudius Gros
for useful discussions. HL, HOJ and RV gratefully acknowledge financial
support from the Deutsche Forschungsgemeinschaft through the
grant FOR 1346. YZ is
supported by National Natural Science Foundation of China
(No. 11174219), Shanghai Pujiang Program (No. 11PJ1409900),
Research Fund for the Doctoral Program of Higher Education of
China (No. 20110072110044) and the Program for Professor of
Special Appointment (Eastern Scholar) at Shanghai Institutions of
Higher Learning.
%%%%%%%%%%%%%%%%%%%%%%%%%%%%%%%%%%%%%%%%%


\begin{thebibliography}{99}
\bibitem{comment1} Within the semiclassical approximation the
auxiliary field is independent of the imaginary time and this hinders
the existence of Fermi liquid behavior.
\bibitem{comment} Note that $\rho(\omega=0)$ and $\beta G(\beta/2)$ become identical as
the temperature approaches zero.
\end{thebibliography}
\end{document}